\documentclass{ws-ijmpd}
\usepackage[super,compress]{cite}
\usepackage{color}

\newcommand{\be}{\begin{equation}}
\newcommand{\ee}{\end{equation}}
\newcommand{\beqa}{\begin{eqnarray}}
\newcommand{\eeqa}{\end{eqnarray}}
\newcommand{\nn}{\nonumber}

\begin{document}

\markboth{M.B. Jahani Poshteh and N. Riazi}
{NED and electrically charged regular BHs}

\catchline{}{}{}{}{}
%
%%%%%%%%%%%%%%%%%%%%%%%%%%%%%%%%%%%%%%%%%%%%%%%%%%%%%%%%%%%%%%%%%%%%

\title{A new nonlinear electrodynamics and electrically charged regular black holes}

\author{Mohammad Bagher Jahani Poshteh}

\address{School of Physics, Institute for Research in Fundamental Sciences (IPM), P.O. Box 19395-5531, Tehran, Iran and\\
	Department of Physics, Institute for Advanced Studies in Basic Sciences (IASBS), Zanjan 45137-66731, Iran\\
jahani@ipm.ir}

\author{Nematollah Riazi}

\address{Department of Physics, Shahid Beheshti University, Tehran 19839, Iran\\
n\underline{\space}riazi@sbu.ac.ir}

\maketitle

\begin{history}
\received{Day Month Year}
\revised{Day Month Year}
\end{history}

\begin{abstract}
A regular static, spherically symmetric electrically charged black hole solution of general relativity coupled to a new theory for nonlinear electrodynamics is presented. This theory has the interesting feature that, at far distances from the black hole, in the weak field limit, the theory reduces to Maxwell Lagrangian with Heisenberg-Euler correction term of quantum electrodynamics. The singular center of the black hole is replaced by flat, de Sitter, or anti de Sitter space, if the spacetime in which the black hole is embedded is asymptotically flat, de Sitter, or anti de Sitter, respectively. Requiring the correspondence to Heisenberg-Euler Lagrangian at large distances, in the weak field limit, we find that (\textit{i}) a minimum mass is required for the formation of an event horizon for the regular static, spherically symmetric solution of the theory, and, (\textit{ii}) the mass of the solution must be quantized. We also study the basic thermodynamic properties of the black hole solution and show that they are qualitatively similar to those of Reissner-Nordstr\"{o}m black hole.
\end{abstract}


\ccode{PACS numbers:}

%\tableofcontents

\section{Introduction}

First examples of black holes in general relativity suffer from singularity at their centers. The singularity theorems of Penrose and Hawking~\cite{penrose1965,hawking1970} prove that, taken the universal applicability of the Einstein theory and under certain energy conditions, singularity is inevitable at the center of black holes, as well as at the big bang~\cite{hawking1967}. The widespread belief is that the singularity shows a disease of the theory and should be resolved~\cite{mtw} (see, however, Ref.~\refcite{horowitz1995} for a different point of view).

Attempts to cure this ``crisis''~\cite{mtw} lead to modification of Einstein equations~\cite{mukhanov1992}. On the other hand, one may couple a special field to Einstein theory to avoid singularity~\cite{frolov1990}. Ay\'{o}n-Beato and Garc\'{\i}a took a first step in this direction by coupling general relativity to nonlinear electrodynamics (NED)~\cite{ayon1998,ayon1999:1,ayon1999:2}. They found exact electrically charged static, spherically symmetric black hole solutions which are regular at the center. The magnetic counterpart of these solutions also appeared in Ref.~\refcite{bronnikov2001}.

The study of NED, initiated by Born and Infeld~\cite{born1934}, was motivated to remove the divergences of self-energy of charged pointlike particles. NED can also resolve the singularity of the electric field at the center of such particles~\cite{kruglov2016}.
NED theories have been coupled to general relativity in different contexts. In Ref.~\refcite{novello1979} the effects of NED on homogeneous isotropic/anisotropic cosmological models have been investigated. The propagation of nonlinear photons in curved spacetime is studied in Refs.~\refcite{drummond1980,alarcon1981}. Also, the black hole solutions in general relativity coupled with NED have been studied in Refs.~\refcite{de1994,soleng1995,palatnik1998}. However, these black hole solutions, which appeared earlier than Ay\'{o}n-Beato and Garc\'{\i}a solution~\cite{ayon1998}, are not regular.

It has been shown in quantum electrodynamics that, one-loop corrections due to the presence of virtual charged particles will result in nonlinear corrections~\cite{heisenberg1936}. In their paper~\cite{heisenberg1936}, Heisenberg and Euler showed that this nonlinear phenomenon would be described with the Lagrangian
\be\label{eqn:helag}
\mathcal{L}_{HE}(\mathcal{F})=\mathcal{F}-\gamma \mathcal{F}^2+\cdots,
\ee
where $\gamma$ is a constant, and $\mathcal{F}=F_{\mu\nu}F^{\mu\nu}/4$, with $F_{\mu\nu}$ denoting the electromagnetic field strength tensor (see also Ref.~\refcite{de1994}).
We should note here that the Heisenberg-Euler Lagrangian is also a function of another electromagnetic invariant $F_{\mu\nu}\tilde{F}^{\mu\nu}$, where $\tilde{F}^{\mu\nu}=\varepsilon^{\mu\nu\alpha\beta}F_{\alpha\beta}/2$ (see for example Ref.~\refcite{ruffini2013}). However, for the case of our interest in this paper (namely, purely electric, static, spherically symmetric case) this term turns out to be identically zero.

Following the seminal papers~\cite{born1934,heisenberg1936}, several alternative theories of NED have been presented in literature~\cite{gaete2014,kruglov2015:1,kruglov2015:2}. Also, in recent decades, some theories of NED have been applied with the Einstein equations to find charged black hole solutions~\cite{hendi2012,yajima2001}.

There is, however, a no-go theorem that forbids, in NED, electrically charged, static, spherically symmetric, general-relativistic black hole solutions with a regular center~\cite{bronnikov1979,bronnikov2000}. The assumption of the theorem is that NED reduces to Maxwell theory in the weak field limit. To overcome the limitations of this no-go theorem, one can use the so-called $\mathcal{P}$ framework of NED proposed by Ay\'{o}n-Beato and Garc\'{\i}a~\cite{ayon1998} (see also Ref.~\refcite{bronnikov2000}). By means of a Legendre transformation, they obtained from the Lagrangian $\mathcal{L}=\mathcal{L}(\mathcal{F})$, the structure function~\cite{salazar1987}:
\be\label{eqn:structure}
\mathcal{H}=2\mathcal{F}\mathcal{L}_{\mathcal{F}}-\mathcal{L},
\ee
where $\mathcal{L}_{\mathcal{F}}=d\mathcal{L}/d\mathcal{F}$. Defining the tensor $P_{\mu\nu}\equiv\mathcal{L}_{\mathcal{F}}F_{\mu\nu}$, one finds that $\mathcal{H}$ is a function of
\be\label{eqn:p}
\mathcal{P}=\frac{1}{4}P_{\mu\nu}P^{\mu\nu}=\mathcal{L}_{\mathcal{F}}^2\mathcal{F},
\ee
because $d\mathcal{H}=\mathcal{L}_{\mathcal{F}}^{-1}d\left(\mathcal{L}_{\mathcal{F}}^2\mathcal{F}\right)=\mathcal{H}_{\mathcal{P}}d\mathcal{P}$, where $\mathcal{H}_{\mathcal{P}}=d\mathcal{H}/d\mathcal{P}=\mathcal{L}_{\mathcal{F}}^{-1}$. Also, the Lagrangian is expressed by~\cite{salazar1987}
\be\label{eqn:lfromh}
\mathcal{L}=2\mathcal{P}\mathcal{H}_{\mathcal{P}}-\mathcal{H}.
\ee

Using this framework, one circumvents the no-go theorem mentioned above, since here there are different Lagrangians in different parts of space and the Lagrangian applicable at the center of the black hole is non-Maxwellian in the weak field regime.

In this paper, we propose a new model of NED. With the NED field as a source in general relativity, we find the regular static, spherically symmetric electrically charged black hole solutions.
Our motivation for presenting this model is, for the part of spacetime at large distances from the black hole, to have a Lagrangian which is compatible with quantum electrodynamics. To this end we propose a model in $\mathcal{P}$ framework which reduces to the Lagrangian (\ref{eqn:helag}) at large distances from the source (the black hole). The NED models which have been proposed in literature to construct {\it electrically} charged black hole solution of general relativity, are compatible with the Lagrangian (\ref{eqn:helag}), at large radial coordinates, only to the leading term.
An interesting feature of our model is that, at large distances from the black hole, it reduces to Maxwell Lagrangian in the first order together with a term proportional to Heisenberg-Euler correction of quantum electrodynamics in the second order.

Compatibility of this model with Heisenberg-Euler Lagrangian, which comes from quantum electrodynamics, results in a number of interesting conclusions. First, the regular static, spherically symmetric solutions of our Einstein-NED theory needs a minimum mass for the formation of an event horizon. And, second, the mass of these solutions is quantized.

We show that weak, dominant, and strong energy conditions are violated near the center of the solution, but at far distances, all of them are satisfied. Also, if the solution has event horizon, it will dress up the region in which the energy conditions are violated.

The outline of this paper is as follows. We present our NED theory in Sec.~\ref{sec:sol}, and by coupling this theory to Einstein equations we find the black hole solutions. Then, considering the curvature invariants, we provide an analysis to the regularity of the solution. In Sec.~\ref{sec:ther} we study the basic thermodynamic behaviors of the black hole solutions. These behaviors are qualitatively similar to those of charged black holes in Einstein-Maxwell theory. The energy conditions are studied in Sec.~\ref{sec:econ}. The compatibility of our NED with Heisenberg-Euler Lagrangian is investigated in Sec.~\ref{sec:qed}. Sec.~\ref{sec:con} is devoted to some concluding remarks. We will work in units where $G = c= k_e = 1$.

\section{Black hole solutions}\label{sec:sol}

The four-dimensional action of general relativity coupled to NED, in the presence of a cosmological constant $\Lambda$, is~\cite{ayon2005} (see Eq.~(\ref{eqn:lfromh}))
\be
\mathcal{S}=\frac{1}{4\pi}\int d^4x\sqrt{-g}\left\{\frac{1}{4}\left(R-2\Lambda\right)-\left[\frac{1}{2}P^{\mu\nu}F_{\mu\nu}-\mathcal{H}(\mathcal{P})\right]\right\},
\ee
in which $g$ is the determinant of the metric tensor and $R$ is the Ricci scalar. Varying this action, one finds the Einstein and electrodynamic field equations
\begin{align}
	G_{\mu}^\nu+\Lambda\delta_{\mu}^\nu&=T_{\mu}^\nu=2\left[\mathcal{H}_{\mathcal{P}}P_{\mu\lambda}P^{\nu\lambda}-\delta_{\mu}^\nu\left(2\mathcal{P}\mathcal{H}_{\mathcal{P}}-\mathcal{H}\right)\right], \label{eqn:einseq}\\
	\nabla_\nu P^{\mu\nu}&=0. \label{eqn:emeq}
\end{align}
where $G_{\mu}^\nu$ is the Einstein tensor and $T_{\mu}^\nu$ is the NED stress-energy tensor.
To solve these equations, we take the static, spherically symmetric metric
\be
ds^2=g_{tt}(r)dt^2+g_{rr}(r)dr^2+r^2d\theta^2+r^2\sin^2\theta d\phi^2,
\ee
and assume the following ansatz for the anti-symmetric tensor
\be\label{eqn:pansatz}
P_{\mu\nu}=2\delta^t_{[\mu}\delta^r_{\nu]}D(r).
\ee
Now, one can show, by using Eq.~(\ref{eqn:einseq}), that $T_t^t=T_r^r$, and
\be
-g_{tt}(r)=\frac{1}{g_{rr}(r)}=f(r).
\ee
Integrating Eq.~(\ref{eqn:emeq}), by using the ansatz (\ref{eqn:pansatz}), we find $D(r)=Q/r^2$, with $Q$ a constant. Then, Eq.~(\ref{eqn:p}) yields $\mathcal{P}=-Q^2/(2r^4)$.

Here, we propose the structure function
\be\label{eqn:myh}
\mathcal{H}(\mathcal{P})=\frac{\mathcal{P}}{(1-\kappa\mathcal{P})^2},
\ee
where $\kappa$ is a free parameter
with the dimension of $({\rm length})^{-2}$.
The motivations for using this $\mathcal{H}$-function are: 1) simplicity, 2) reduction to the Maxwell theory at large distances (weak field limit), 3) avoidance of spacetime singularity, and most innovatively 4) accordance with Heisenberg-Euler theory of quantum electrodynamics at large distances from the black hole, in the weak field limit (as will be cleared later in the paper).
Since $\mathcal{P}$ is a negative quantity we see that $\kappa$ should be positive for the structure function (\ref{eqn:myh}) not to have a divergent point.

\begin{figure}[htp]
	\centering
	\includegraphics[width=0.45\textwidth]{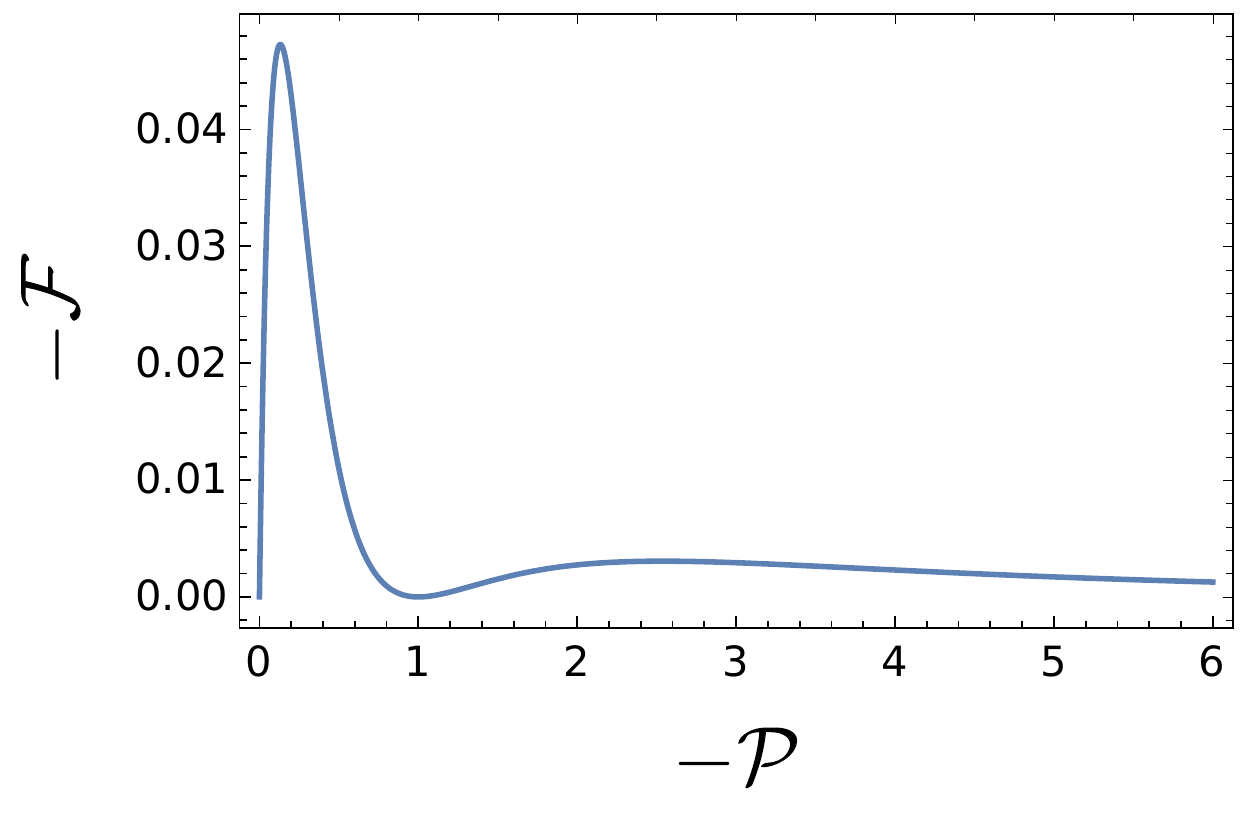}
	\includegraphics[width=0.45\textwidth]{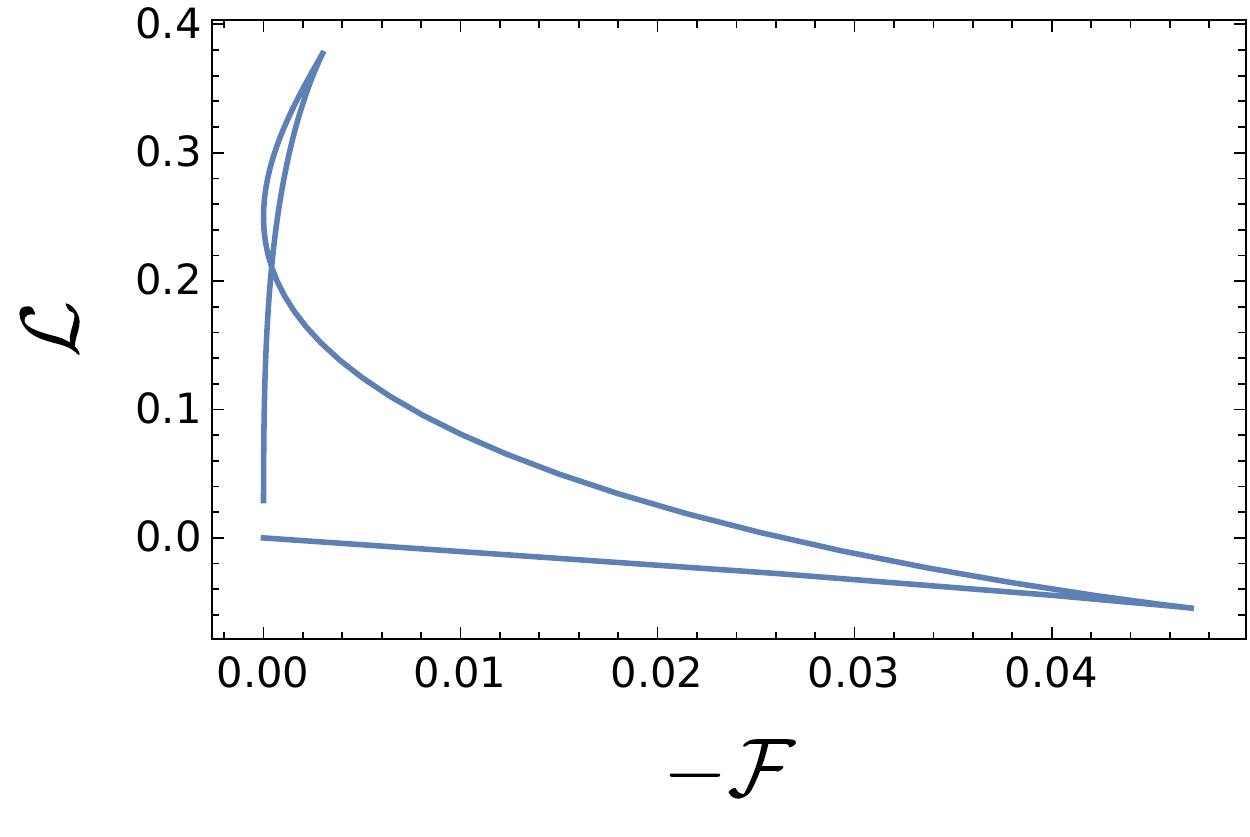}
	\caption{\textit{Left}: We see that the function $\mathcal{F}(\mathcal{P})$ has three ectrema (four monotonicity ranges). \textit{Right}: $\mathcal{L}(\mathcal{F})$ has been presented. There are four Lagrangians in the spacetime corresponding to the four monotonicity ranges of $\mathcal{F}(\mathcal{P})$. We have taken $\kappa=1\,{\rm m}^{-2}$ in both plots.}
	\label{fig:Lag_formalism}
\end{figure}

The Lagrangian dynamics is, however, specified in the $\mathcal{F}$ framework; rendering the $\mathcal{P}$ framework secondary. As it has been pointed out in Ref.~\refcite{bronnikov2001}, $\mathcal{F}$ and $\mathcal{P}$ frameworks are only equivalent where $\mathcal{F}$ is a monotonic function of $\mathcal{P}$. By using Eqs.~(\ref{eqn:p}) and (\ref{eqn:myh}) we can find that
\be
\mathcal{F}=\mathcal{H}_\mathcal{P}^2 \mathcal{P}=\frac{P (\kappa P+1)^2}{(\kappa P-1)^6}.
\ee
In the left panel of Fig.~\ref{fig:Lag_formalism} we have presented $\mathcal{F}$ as a function of $\mathcal{P}$. $\mathcal{F}(\mathcal{P})$ has three ectrema, so there are four monotonicity ranges through the spacetime. Each of these monotonicity ranges correspond to a Lagrangian as a function of $\mathcal{F}$. We have presented these Lagrangians in the right panel of Fig.~\ref{fig:Lag_formalism}. It is obvious that $\mathcal{L}(\mathcal{F})$ suffers branching and there are four Lagrangians. We will show in Sec.~\ref{sec:qed} that the Lagrangian in large distances from the source is consistent with the Heisenberg-Euler Lagrangian of quantum electrodynamics. We will also find the Lagrangian for small radial coordinate.

\begin{figure}[htp]
	\centering
	\includegraphics[width=0.45\textwidth]{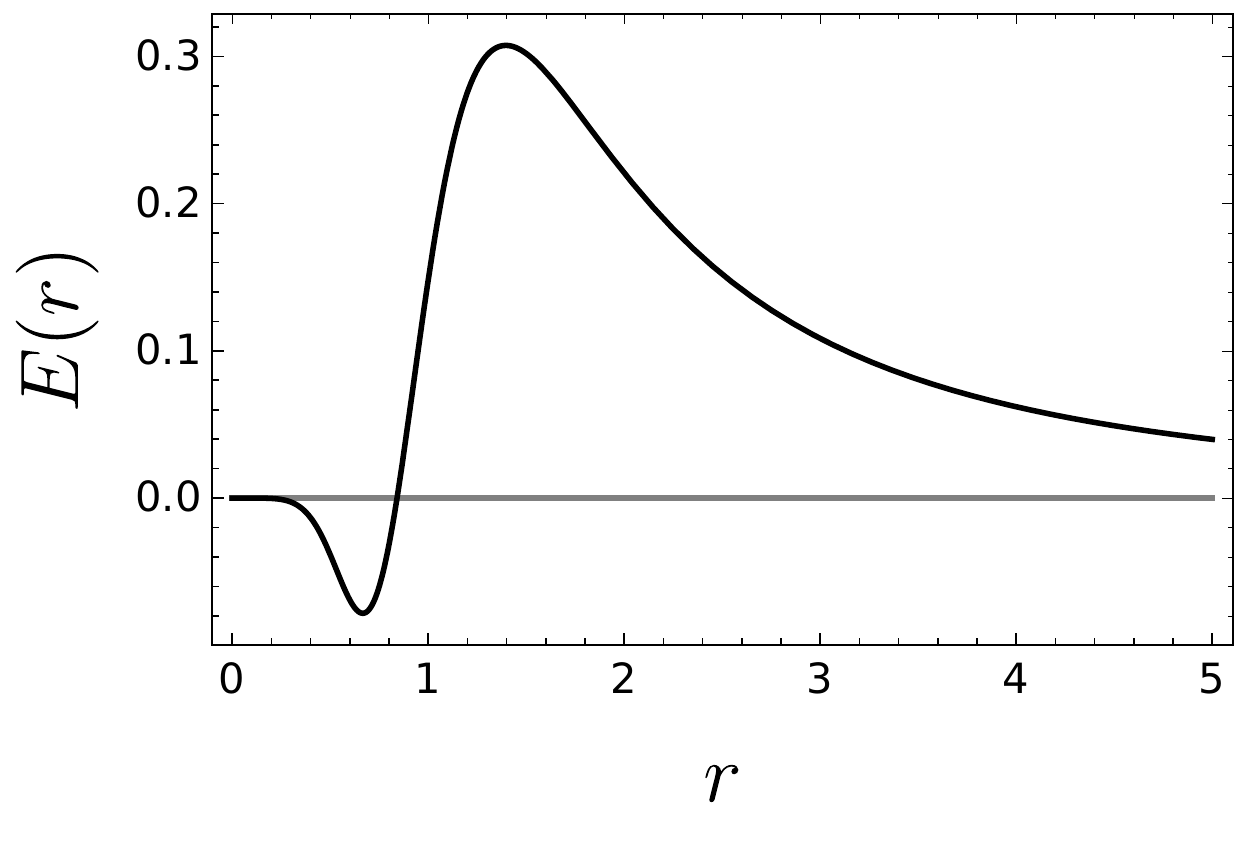}
	\caption{The electric field as a function of radius. We see that the electric field is well behaved at the center as well as elsewhere. We have taken $\kappa=1\,{\rm m}^{-2}$ and $Q=1\,{\rm m}$.}
	\label{fig:elec_field}
\end{figure}

The equation for the electric field is
\be\label{eqn:e}
E(r)=F_{tr}=\mathcal{H}_{\mathcal{P}}P_{tr}=\frac{4 Q r^6 \left(2 r^4-\kappa Q^2\right)}{\left(2 r^4+\kappa Q^2\right)^3},
\ee
with the asymptotic behavior
\be
E(r\rightarrow \infty)\approx \frac{Q}{r^2}+\mathcal{O}\left(\frac{1}{r^3}\right).
\ee
Thus, we regard $Q$ as the electric charge. It is obvious from Eq.~(\ref{eqn:e}) that the electric field approaches zero as $r\rightarrow 0$. It has also another zero at $r=\sqrt[4]{\kappa Q^2/2}$, and two extrema at
\be
r=\sqrt[4]{\frac{1}{2}\left(4\mp\sqrt{13}\right)\kappa Q^2}.
\ee
We have plotted the electric field (\ref{eqn:e}) in Fig.~\ref{fig:elec_field}.

Substituting the metric function
\be\label{eqn:f}
f(r)=1-\frac{2\mathcal{M}(r)}{r},
\ee
into the field equations (\ref{eqn:einseq}), we find from the $^t_t$ component (other components give no new information):
\be
-2\frac{d\mathcal{M}(r)}{dr}=r^2\left(2\mathcal{H}-\Lambda\right).
\ee
Integrating this equation we find
\be\label{eqn:mym}
\mathcal{M}(r)=B(r)+\frac{\Lambda r^3}{6},
\ee
with
\beqa\label{eqn:myb}
B(r)&=&-\frac{3}{16}\sqrt[4]{\frac{Q^6}{2\kappa}}\left[\tanh ^{-1}\left(\frac{\sqrt[4]{8\kappa Q^2}r}{\sqrt{\kappa Q^2}+\sqrt{2} r^2}\right)+\tan ^{-1}\left(1-\sqrt[4]{\frac{8}{\kappa Q^2}}r\right)\right.\nn\\
&-&\left.\tan ^{-1}\left(1+\sqrt[4]{\frac{8}{\kappa Q^2}}r\right)\right]- \frac{Q^2 r^3}{4 \kappa Q^2+8 r^4}+c_0,
\eeqa
in which $c_0$ is the constant of integration.

The total mass $M$ of the black hole is found via $M=B(r\rightarrow \infty)$.
\footnote{It will be shown in Sec.~\ref{sec:econ} that $M$ is the ADM mass.}
Also, we require the metric function $f(r)$ to be non-singular at $r=0$. This would be satisfied if
\be\label{eqn:freepara}
\kappa =\frac{81 \pi ^4 Q^6}{131072 M^4}, \qquad c_0=0.
\ee

Using Eqs.~(\ref{eqn:mym}) and (\ref{eqn:freepara}), we find asymptotic and near center approximations of the metric function, respectively, as
\be\label{eqn:f:asymp}
f_{{\rm asymp}}\approx 1-\frac{2 M}{r}+\frac{Q^2}{r^2}-\frac{\Lambda}{3}r^2, \qquad
f_{{\rm nc}}\approx 1-\frac{\Lambda}{3}r^2.
\ee
We see that the solution asymptotically behaves like Reissner-Nordstr\"{o}m (anti) de Sitter black hole. On the other hand, one finds from the near horizon expansion that, the singular center is replaced by a flat, de Sitter, or anti de Sitter core for zero, positive, or negative $\Lambda$, respectively.

To see if the singularity at the center is really resolved, we study the curvature invariants. The explicit expressions of the curvature invariants $R$, $R_{\alpha\beta}R^{\alpha\beta}$, and $R_{\alpha\beta\gamma\delta}R^{\alpha\beta\gamma\delta}$ are too lengthy, however, their near center behaviors are
\beqa\label{eqn:curinvs}
\lim\limits_{r\rightarrow 0} R &=& 4\Lambda, \nn\\
\lim\limits_{r\rightarrow 0} R_{\alpha\beta}R^{\alpha\beta} &=& 4\Lambda^2,\\
\lim\limits_{r\rightarrow 0} R_{\alpha\beta\gamma\delta}R^{\alpha\beta\gamma\delta} &=& \frac{8 \Lambda^2}{3}. \nn
\eeqa
Therefore, the curvature invariants do not diverge at the center.
\footnote{For the sake of comparison, we recall that for Reissner-Nordstr\"{o}m (anti) de Sitter black hole, as $r\rightarrow 0$, we have $R\rightarrow 4\Lambda$, $R_{\alpha\beta}R^{\alpha\beta}\rightarrow 4Q^4/r^8$, and $ R_{\alpha\beta\gamma\delta}R^{\alpha\beta\gamma\delta}\rightarrow 56Q^4/r^8$, which show a singularity at the center.}
Neither they diverge in any region of spacetime. It is known that in an arbitrary spacetime there are fourteen algebraically independent curvature invariants which contain, at most, second order derivative of the metric~\cite{thomas1934,lake1994}. To make sure that all such invariants are regular we follow the instruction of Ref.~\refcite{fan2016} (see also Ref.~\refcite{torres2017}). The function $\mathcal{M}(r)$ \eqref{eqn:mym} must be a smooth function which is at least three times differentiable (a $C^3$ function). It is also required that $\mathcal{M}(r)$, $\mathcal{M}'(r)$, and $\mathcal{M}''(r)$ (where a prime denotes differentiation with respect to $r$) vanish at the center and $\mathcal{M}'''(r)$ approaches a finite value (not necessarily zero) as $r\rightarrow 0$. Indeed, one can easily check that
\beqa
\lim\limits_{r\rightarrow 0} \mathcal{M}(r) &=& 0, \qquad
\lim\limits_{r\rightarrow 0} \mathcal{M}'(r) = 0, \nn\\
\lim\limits_{r\rightarrow 0} \mathcal{M}''(r) &=& 0, \qquad
\lim\limits_{r\rightarrow 0} \mathcal{M}'''(r) = \Lambda,
\eeqa
Then the sufficient conditions for the regularity of curvature invariants is that $\mathcal{M}(r)/r^3$ is finite in the $r\rightarrow 0$ limit. Using Eq.~\eqref{eqn:mym} we find that
\be
\lim\limits_{r\rightarrow 0} \frac{\mathcal{M}(r)}{r^3} = \frac{\Lambda }{6}-\frac{32768 M^4}{81 \pi ^4 Q^6}.
\ee
Therefore, as long as $Q\neq 0$, the spacetime regularity is verified.

\begin{figure}[htp]
	\centering
	\includegraphics[width=0.45\textwidth]{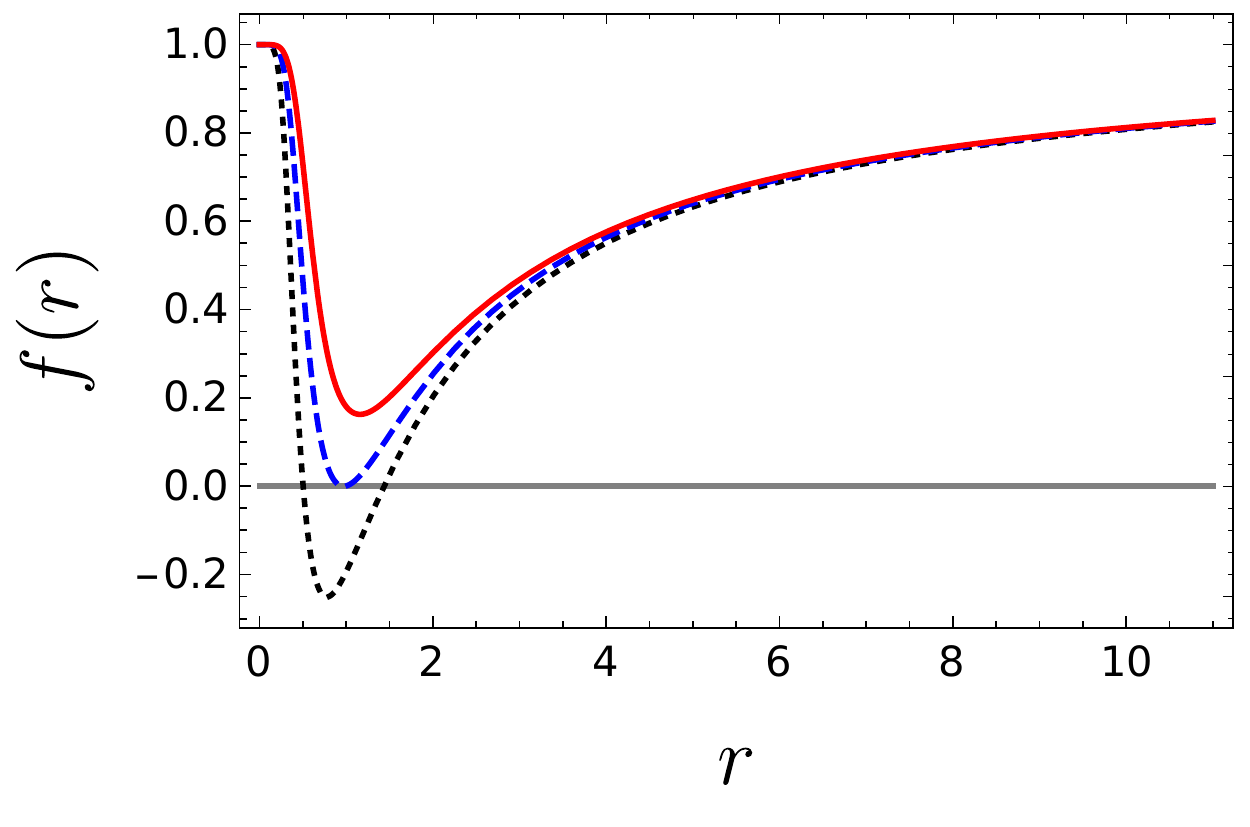}
	\includegraphics[width=0.45\textwidth]{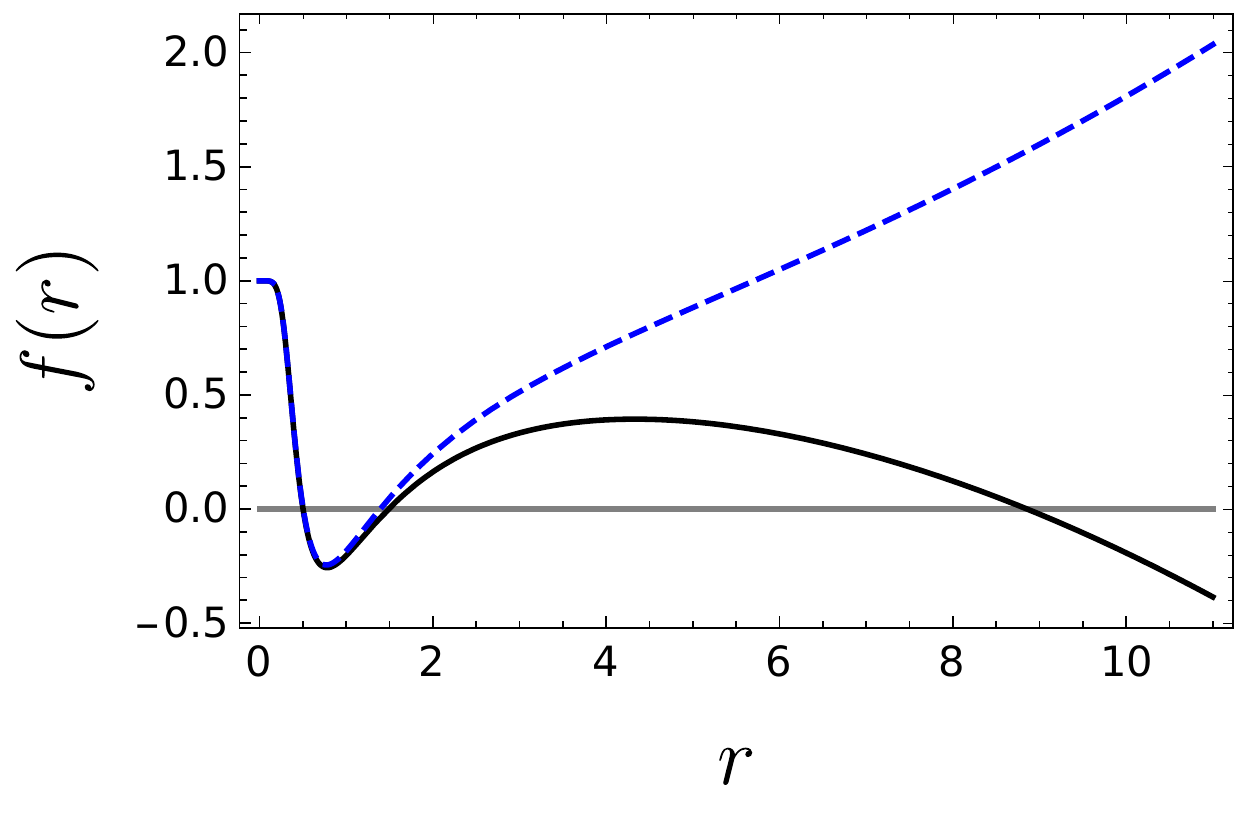}
	\caption{\textit{Left}: The metric function of our static, spherically symmetric Einstein-NED solution in asymptotically flat spacetime. We have fixed $M=1\,{\rm m}$, and $Q=0.9\,{\rm m}$ (dotted, black curve), $Q=1.00658\,{\rm m}$ (dashed, blue curve), or $Q=1.1\,{\rm m}$ (solid, red curve). \textit{Right}: The metric function in the presence of a cosmological constant. Here, $M=1\,{\rm m}$, $Q=0.9\,{\rm m}$, and $\Lambda=0.03\,{\rm m}^{-2}$ (solid, black curve) or $\Lambda=-0.03\,{\rm m}^{-2}$ (dashed, blue curve).}
	\label{fig:f}
\end{figure}

In Fig.~\ref{fig:f}, we have plotted the metric function in asymptotically flat and (anti) de Sitter spacetimes. Depending on the values of $M$, $Q$, and $\Lambda$, the metric function may have none, one, two, or three zeros. By noting that the curvature invariants are non-singular throughout spacetime, $f(r)=0$ (or, equivalently, the divergence of $g_{rr}$) denotes a coordinate singularity and is associated with a horizon. In the case of asymptotically flat spacetime, one may have a (extremal) black hole with an event horizon, or a black hole with both Cauchy and event horizons. Using numerical methods we find that, for asymptotically flat spacetimes, the event horizon exists if $|Q|\leq 1.00658\,M$. Event and Cauchy horizons may also form in asymptotically (anti) de Sitter spacetimes. For de Sitter solutions (with a positive $\Lambda$) there is a cosmological horizon, as well. This one is present whether or not one has an event (and Cauchy) horizon.

We close this section by considering the $Q\rightarrow 0$ limit of our black hole solution. Using Eqs.~\eqref{eqn:f}, \eqref{eqn:mym}, and \eqref{eqn:myb} we find that
\be\label{eqn:f:sch}
\lim\limits_{Q\rightarrow 0} f(r)=1-\frac{2 M}{r}-\frac{\Lambda  r^2}{3},
\ee
which is the metric function of Schwarzschild (anti) de Sitter black hole. Also, in this limit, the curvature invariants $R$, $R_{\alpha\beta}R^{\alpha\beta}$, and $R_{\alpha\beta\gamma\delta}R^{\alpha\beta\gamma\delta}$ reduce to
\beqa\label{eqn:curinvs:shc}
\lim\limits_{Q\rightarrow 0} R &=& 4\Lambda, \nn\\
\lim\limits_{Q\rightarrow 0} R_{\alpha\beta}R^{\alpha\beta} &=& 4\Lambda^2, \\
\lim\limits_{Q\rightarrow 0} R_{\alpha\beta\gamma\delta}R^{\alpha\beta\gamma\delta} &=& \frac{48 M^2}{r^6} + \frac{8 \Lambda^2}{3}. \nn
\eeqa
One sees that the Kretschmann scalar diverges at $r = 0$, indicating a physical singularity at the center of the black hole. Hence, we conclude that the regularity of the solution \eqref{eqn:f} is a result of the NED \eqref{eqn:myh} which is coupled to the Einstein equations.

\section{Thermodynamics}\label{sec:ther}

Since the solution presented in the previous section could have an event horizon, one can associate a thermodynamic to it. In this section we briefly study the basic thermodynamic properties of the solution. The metric function \eqref{eqn:f} of the static, spherically symmetric black hole solution has a complicated form (see Eqs.~\eqref{eqn:mym}, \eqref{eqn:myb}, and \eqref{eqn:freepara}). Therefore, we are not able to solve $f(r_H)=0$ to find an analytic expression of the mass $M$ as a function of horizon radius $r_H$. However, we can use numeric technique to study the basic thermodynamic quantity of the black hole.

\begin{figure}[htp]
	\centering
	\includegraphics[width=0.4\textwidth]{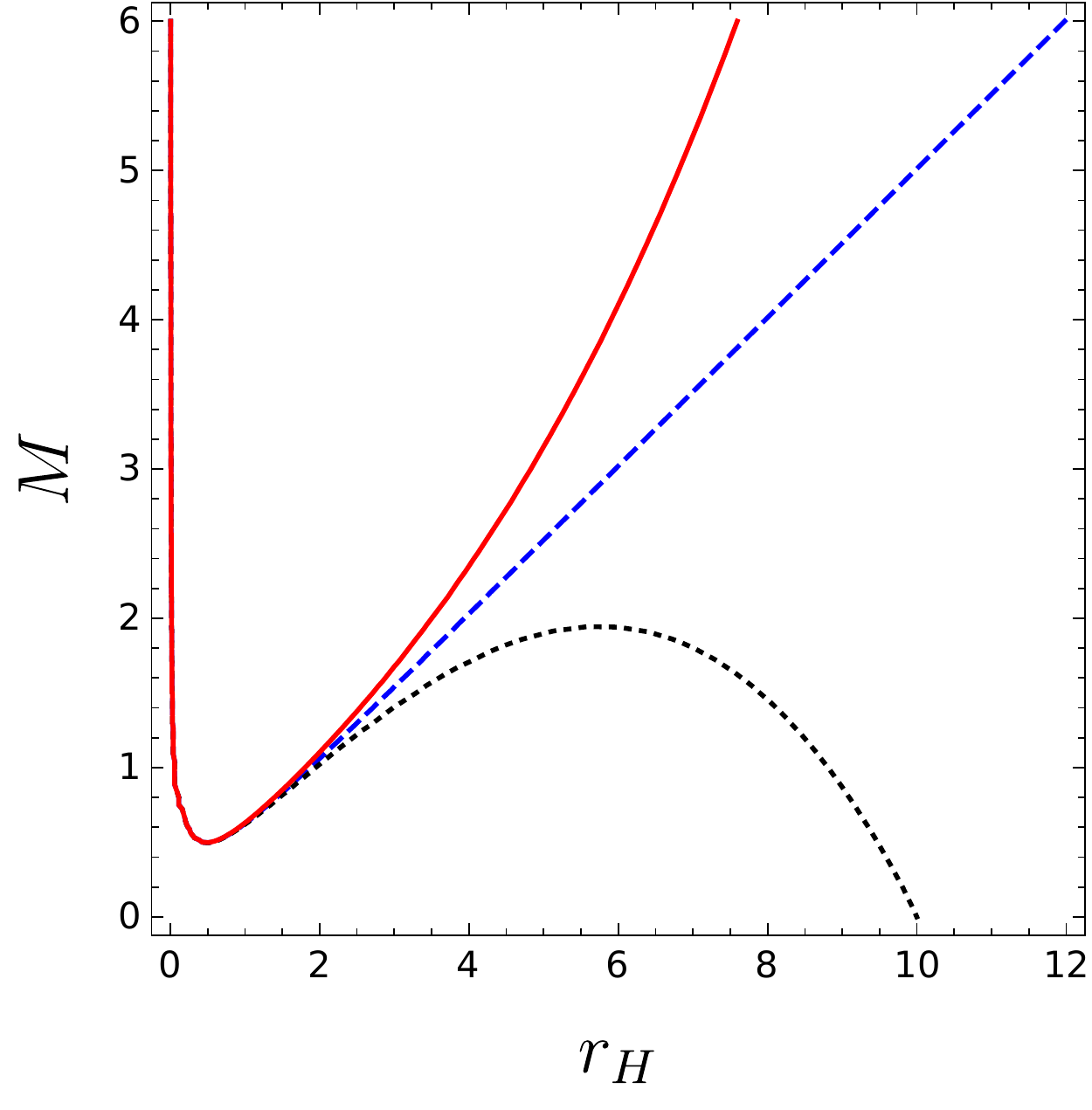}
	\caption{The total mass as a function of the horizon radius. We have taken $Q=0.5\,{\rm m}$ and $\Lambda=-0.03\,{\rm m}^{-2}$ (solid, red curve), $\Lambda=0$ (dashed, blue curve), and $\Lambda=0.03\,{\rm m}^{-2}$ (dotted, black curve).}
	\label{fig:mass}
\end{figure}

In Fig.~\ref{fig:mass} we have plotted the total mass as a function of the horizon radius for flat, de Sitter, and anti de Sitter spaces. The mass has a minimum in all three cases. The radius below the radius at which the mass is minimum corresponds to inner horizon radius. For the cases of asymptotically flat and asymptotically anti de Sitter black holes, the radius on the right hand side of minimum point corresponds to event horizon radius. For the case of asymptotically de Sitter black holes, the mass has a maximum. In this case the radius between the minimum and maximum corresponds to event horizon radius, and the radius on the right hand side of the maximum corresponds to cosmological horizon radius. Qualitatively similar behaviors can be shown to be the case for the mass of Reissner-Nordstr\"{o}m black holes by using Eq.~\eqref{eqn:f:asymp}.

\begin{figure}[htp]
	\centering
	\includegraphics[width=0.45\textwidth]{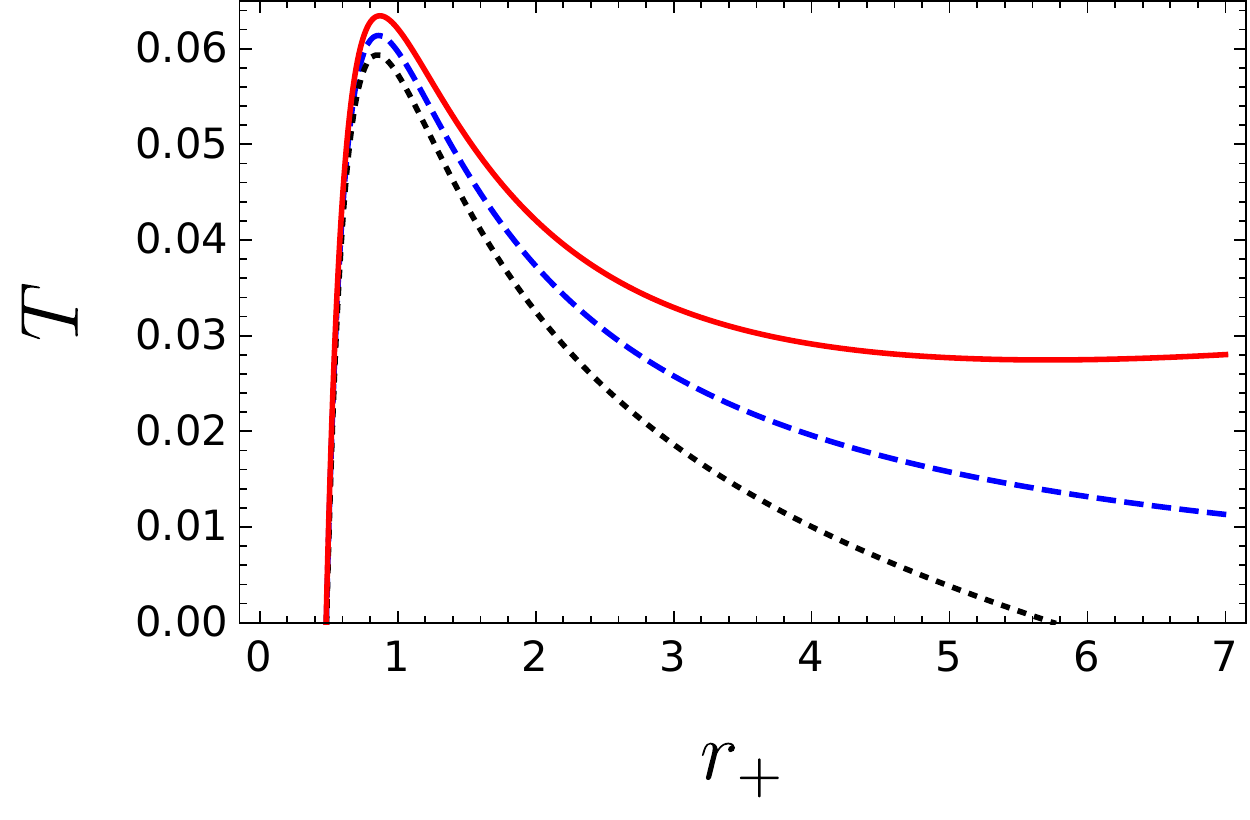}
	\includegraphics[width=0.45\textwidth]{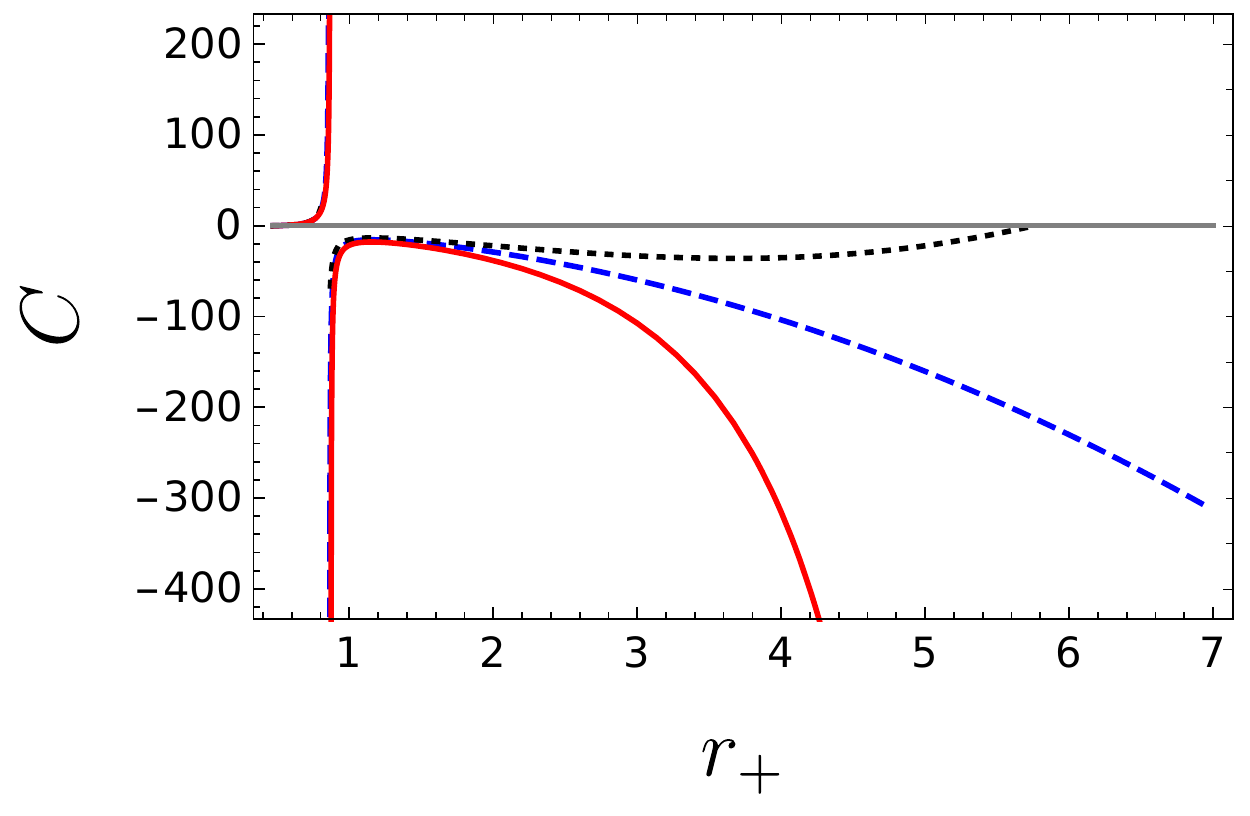}
	\caption{\textit{Left}: The temperature as a function of the event horizon radius. \textit{Right}: The specific heat as a function of the event horizon radius. We have taken $Q=0.5\,{\rm m}$ and $\Lambda=-0.03\,{\rm m}^{-2}$ (solid, red curve), $\Lambda=0$ (dashed, blue curve), and $\Lambda=0.03\,{\rm m}^{-2}$ (dotted, black curve) in both figures.}
	\label{fig:tem:c}
\end{figure}

The temperature associated to the event horizon is defined as
\be
T=\frac{f'(r=r_+)}{4\pi}.
\ee
We take the entropy of the black hole to be given by Bekenstein-Hawking entropy as one quarter of the event horizon area~\cite{bekenstein1973,hawking1975}, so $S=\pi r_+^2$. The specific heat can then be find through
\be
C=T\frac{\partial S}{\partial T}.
\ee
In the left panel of Fig.~\ref{fig:tem:c} we have plotted the temperature as a function of the event horizon radius. We see that the temperature is always non-negative as long as we are considering the horizon radius which corresponds to the \textit{event} horizon radius. The temperature is zero for the event horizon radius that equals the inner horizon radius (which occurs at the local minimum of Fig.~\ref{fig:mass}), and, also, for the case of asymptotically de Sitter black holes, for the event horizon radius that equals the cosmological horizon radius (which occurs at the local maximum of Fig.~\ref{fig:mass}). Similar behaviors appear for the Reissner-Nordstr\"{o}m black holes.

The temperature has a maximum at which the specific heat diverges. This is shown in the right panel of Fig.~\ref{fig:tem:c}. The specific heat is positive for small black holes which means that these black holes are locally stable. For the large black holes, on the right hand side of the divergent point, the specific heat is negative which indicates that these black holes are locally unstable. However, the case in anti de Sitter background needs more considerations.

\begin{figure}[htp]
	\centering
	\includegraphics[width=0.45\textwidth]{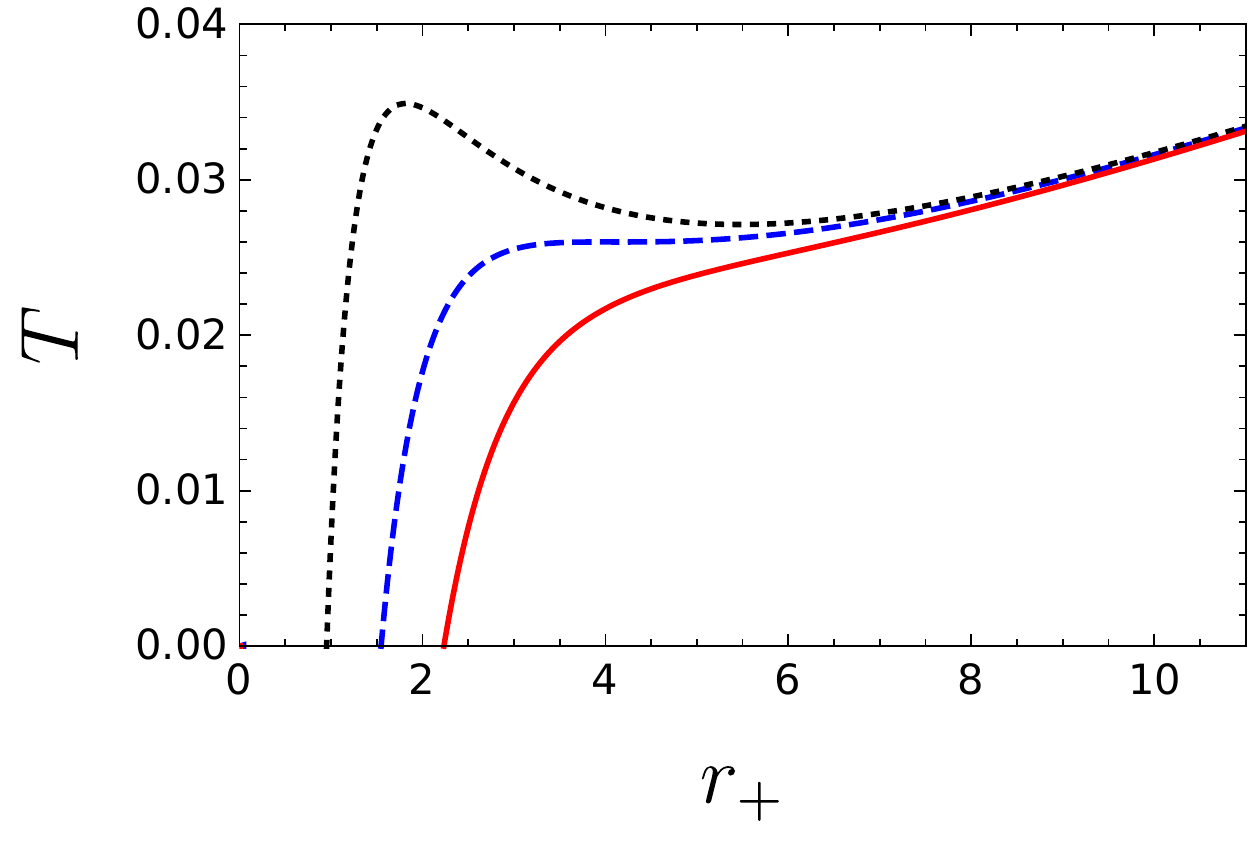}
	\includegraphics[width=0.45\textwidth]{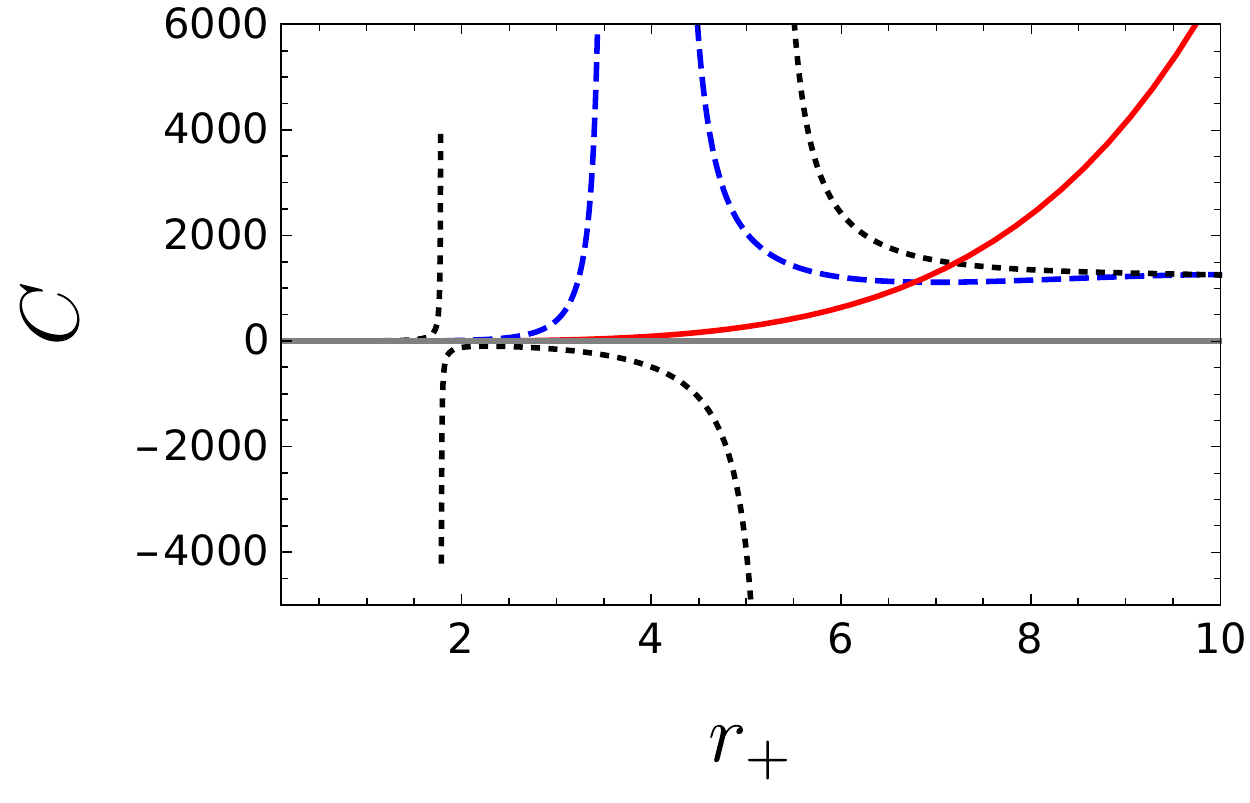}
	\caption{\textit{Left}: The temperature as a function of event horizon radius in anti de Sitter background. \textit{Right}: The specific heat as a function of event horizon radius in anti de Sitter background. We have taken $\Lambda=-0.03\,{\rm m}^{-2}$ and $Q=2.5\,{\rm m}>Q_{crit}$ (solid, red curve), $Q=Q_{crit}=1.66504\,{\rm m}$ (dashed, blue curve), and $Q=1\,{\rm m}<Q_{crit}$ (dotted, black curve) in both figures.}
	\label{fig:tem:c:ads}
\end{figure}

As for the black holes in Einstein-Maxwell theory~\cite{chamblin1999:1,chamblin1999:2,kubizvnak2012}, the thermodynamics of black holes in anti de Sitter background is more interesting --- it is not specific to Einstein theory; see for example Refs.~\refcite{poshteh2017,poshteh2021}. In Fig.~\ref{fig:tem:c:ads} we have plotted the temperature and specific heat for a fixed value of negative cosmological constant and different charges $Q$.

For those charges whose absolute values are smaller than a critical value $|Q_{crit}|$, the temperature has two extrema at which the specific heat diverges. The specific heat is negative between these divergent points and positive elsewhere. This means that for black holes with $|Q|<|Q_{crit}|$, the intermediate size/mass black hole is locally unstable but small and large black holes are locally stable.

For black holes with $|Q|=|Q_{crit}|$ one sees an inflection point in the temperature. The specific heat blows up at the inflection point and is positive everywhere. For black holes with $|Q|>|Q_{crit}|$ there is no extremum in the temperature. In this case there is no divergent point in the specific heat for finite values of the event horizon radius. Such black holes are always in a locally stable state.

Indeed, these behaviors are similar to those of Reissner-Nordstr\"{o}m-anti de Sitter black holes~\cite{chamblin1999:1,chamblin1999:2}. One can find that in Einstein-Maxwell theory $Q_{crit,{\rm RN}}^2=-1/(12\Lambda)$~\cite{chamblin1999:1}. Here we are not able to find an analytic expression for $Q_{crit}$, but the numerical calculations that we have used in Fig.~\ref{fig:tem:c:ads} show that, for a fixed value of the cosmological constant, $|Q_{crit}|<|Q_{crit,{\rm RN}}|$. (In fact, for the values used in Fig.~\ref{fig:tem:c:ads} the difference between $|Q_{crit}|$ and $|Q_{crit,{\rm RN}}|$ is about 1 part in 1000.)

\section{Energy conditions}\label{sec:econ}

Here we study the energy conditions for the black hole spacetime presented in Sec.~\ref{sec:sol}. By using Eqs.~(\ref{eqn:einseq}) and (\ref{eqn:myh}), one can easily find the energy density $\rho$, radial pressure $p_r$, and tangential pressure $p_\theta$ and $p_\phi$ as
\begin{align}\label{eqn:rhop}
	p_r&=T^r_r=T^t_t=-\rho=2\mathcal{H}=-\frac{4 Q^2 r^4}{\left(\kappa Q^2+2 r^4\right)^2}, \nn\\
	p_\theta&=T^\theta_\theta=T^\phi_\phi=p_\phi=2\mathcal{H}-4\mathcal{P}\mathcal{H}_{\mathcal{P}}=\frac{8Q^2r^8-12\kappa Q^4r^4}{\left(\kappa Q^2+2 r^4\right)^3}.
\end{align}

For the metric to be non-singular at the center, $\kappa$ would be given by Eq.~(\ref{eqn:freepara}), which is positive. This guarantees the non-singularity of the structure function (\ref{eqn:myh}). From Eq.~(\ref{eqn:rhop}), one sees that for a positive value of $\kappa$, the energy density and the components of pressure are also regular in all regions of spacetime.
We would also like to note that, using the energy density \eqref{eqn:rhop} along with the definition of ADM mass, one can show that
\be
M_{ADM}=\frac{1}{2}\int_{0}^{\infty}r^2\rho(r)dr=M,
\ee
so the mass parameter $M$ is indeed the ADM mass.

Weak Energy Condition (WEC) is satisfied if~\cite{hawking1973,dymnikova2004}
\be
\rho\geq 0, \qquad \rho+p_k\geq 0,
\ee
with $k=r, \theta, \phi$. Dominant Energy Condition (DEC) requires~\cite{hawking1973,dymnikova2004}
\be
\rho\geq 0, \qquad \rho+p_k\geq 0, \qquad \rho-p_k\geq 0.
\ee
Using Eq.~(\ref{eqn:rhop}), one can show that WEC and DEC are both satisfied for $\mathcal{P}\geq -1/\kappa$, or equivalently for $r\geq r_{{\rm WEC}}=\sqrt[4]{\kappa Q^2/2}$.

Strong Energy Condition (SEC) requires~\cite{hawking1973,dymnikova2004}
\be
\rho+\sum p_k\geq 0.
\ee
It is easy to show that this condition holds if $\mathcal{P}\geq -1/(3\kappa)$, or, equivalently, $r\geq r_{{\rm SEC}} = \sqrt[4]{3\kappa Q^2/2}$. We see that $r_{{\rm SEC}}>r_{{\rm WEC}}$, so wherever SEC is satisfied so will be WEC and DEC.

\begin{figure}[htp]
	\centering
	\includegraphics[width=0.45\textwidth]{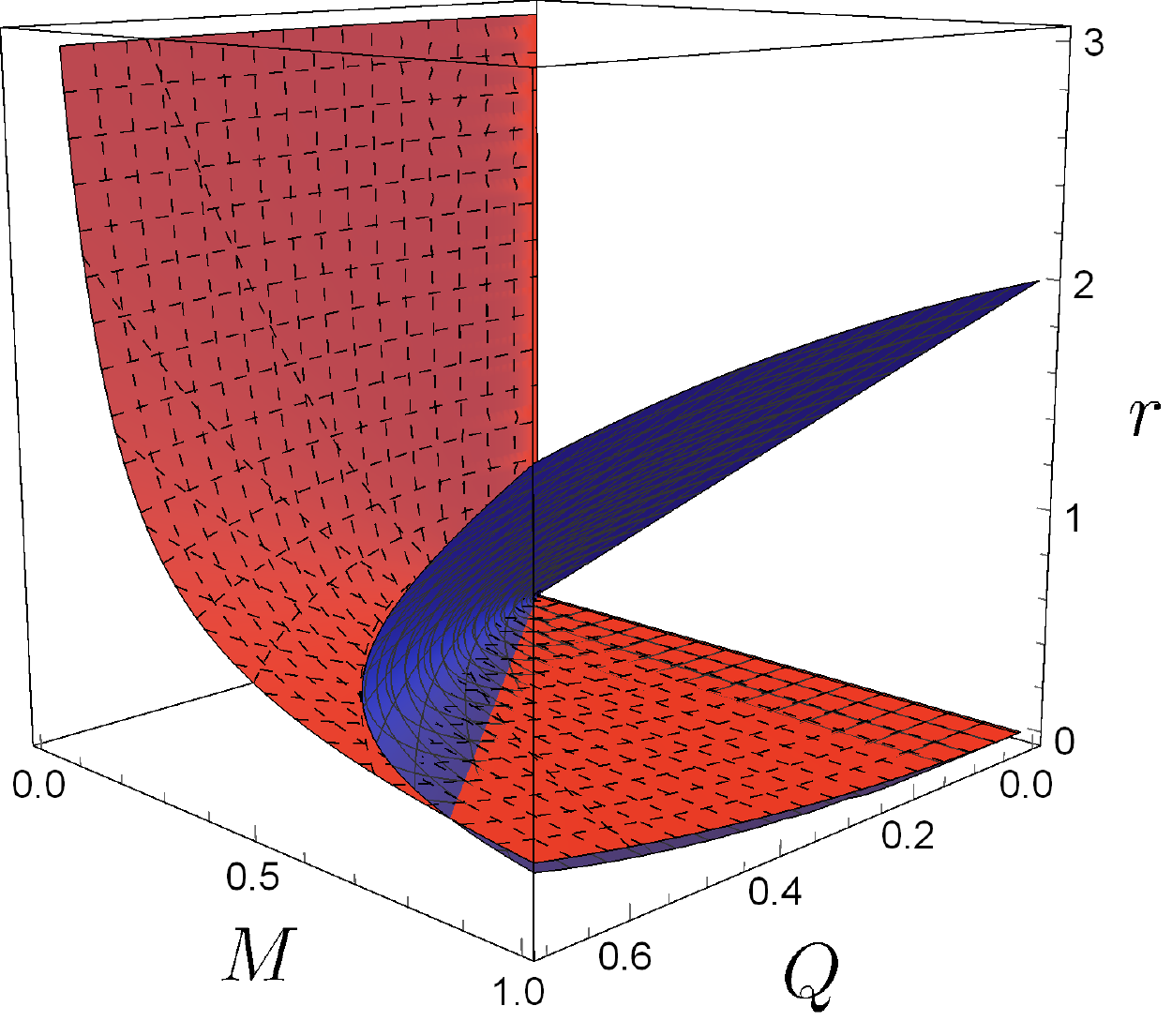}
	\caption{The red surface (which has been meshed with dashed curves) represent the $r=r_{{\rm SEC}}$. For any points above this surface WEC, DEC, and SEC are satisfied. The blue surface represents the points at which $f(r)=0$. The top part of this surface (with larger $r$) corresponds to the event horizon. The bottom part of the blue surface (with smaller $r$) corresponds to the inner horizon. Wherever in the $M-Q$ plane the event horizon exists, its radius is larger than $r_{{\rm SEC}}$.}
	\label{fig:shielding}
\end{figure}

We have solved $f(r)=0$ perturbatively to find an approximate relation for the radius of the event horizon $r_+$. The relation is too involved to write here, but, we have been able to show that $r_+\geq r_{{\rm SEC}}$. Therefore, if the event horizon exists, it would dress up the region of WEC/DEC/SEC violation. Outside the event horizon, all of these energy conditions are satisfied.
\footnote{For comparison with Reissner-Nordstr\"{o}m black hole we note that in that case one has $\mathcal{H}=\mathcal{P}$, so, by using Eq.~(\ref{eqn:rhop}), one can show that WEC, DEC, and SEC are satisfied everywhere in the spacetime.}
This point has been shown in Fig.~\ref{fig:shielding}. We see that, wherever in the $M-Q$ plane the event horizon exists, it would shield the region in which WEC, DEC, and SEC are violated. One finds that the inner horizon may or may not shield this region. Therefore, depending on the values of the mass and the electric charge of the black hole, there might be a region between the inner and event horizon in which the energy conditions are violated.

As a final remark of this section we note that in the limit $Q\rightarrow 0$ both $r_{{\rm SEC}}$ and $r_{{\rm WEC}}$ vanish, which means that in this limit the energy conditions WEC, DEC, and SEC are satisfied everywhere in the spacetime. This has been expected from Eq.~\eqref{eqn:f:sch} since in the $Q\rightarrow 0$ limit our solution reduces to Schwarzschild (anti) de Sitter black hole which has a singularity at its center (see Eqs.~\eqref{eqn:curinvs:shc}).

\section{Quantum electrodynamic considerations}\label{sec:qed}

For the electrically charged, static, spherically symmetric solutions that we are considering in this paper, the Heisenberg-Euler Lagrangian is given by Eq.~(\ref{eqn:helag}).
Considering this Lagrangian at far distances from the black hole, and using Eq.~(\ref{eqn:structure}), one finds the structure function associated with the Heisenberg-Euler Lagrangian in terms of $\mathcal{F}$. In the weak field limit ($\mathcal{F}\rightarrow 0$), it is
\be\label{eqn:h(f)}
\mathcal{H}(\mathcal{F})=\mathcal{F}-3 \gamma \mathcal{F}^2.
\ee
Also, from Eq.~(\ref{eqn:p}) we have, to second order in $\mathcal{F}$
\be
\mathcal{P}=\mathcal{F}-4 \gamma \mathcal{F}^2.
\ee
This equation can be solved for $\mathcal{F}$. Noticing that in the Maxwell limit $\mathcal{L}(\mathcal{F})=\mathcal{F}=\mathcal{P}=\mathcal{H}(\mathcal{P})$, one finds
\be
\mathcal{F}=\frac{1 -\sqrt{1-16 \gamma \mathcal{P}}}{8 \gamma}.
\ee
Substituting this equation into Eq.~(\ref{eqn:h(f)}), we find, for small $\mathcal{P}$,
\be\label{eqn:h(p)}
\mathcal{H}(\mathcal{P})\approx \mathcal{P}+ \gamma \mathcal{P}^2 + \cdots.
\ee

Therefore, we find that, for the structure function to be compatible with Heisenberg-Euler Lagrangian (\ref{eqn:helag}) of quantum electrodynamics, it must have an expansion of the form (\ref{eqn:h(p)}). The structure functions which have been presented in Refs.~\refcite{ayon1998,ayon1999:1,ayon1999:2,ayon2005} satisfy the plausible condition of correspondence to Maxwell theory $\mathcal{H}(\mathcal{P}\rightarrow 0)\approx \mathcal{P}$, but the second term in their expansion is either proportional to $(-\mathcal{P})^{5/4}$~\cite{ayon1998,ayon1999:2,ayon2005}, or $(-\mathcal{P})^{3/2}$~\cite{ayon1999:1}.

Since $\mathcal{P}$ approaches zero at large distances from the black hole, the asymptotic expansion of the structure function (\ref{eqn:myh}) is
\be\label{eqn:myhserie}
\mathcal{H}(\mathcal{P})\approx \mathcal{P}+2\kappa \mathcal{P}^2.
\ee
We demand that, at far distances from the black hole, the structure function (\ref{eqn:myh}) should be consistent with the Heisenberg-Euler Lagrangian in the weak field limit. Then, comparing Eqs.~(\ref{eqn:myhserie}) and (\ref{eqn:h(p)}), we find $2\kappa=\gamma$. The coupling constant $\gamma$ has been found in the weak field approximation of quantum electrodynamics to be~\cite{heisenberg1936,yajima2001}
\be
\gamma=\frac{2e^6}{45\pi \alpha m_e^4},
\ee
in which $\alpha$ denotes the fine structure constant and $e$ and $m_e$ are the charge and the mass of electron, respectively.
The electron is invoked in the theory as a member of virtual electron-positron pair produced by photon propagators. It then follows that, for the black hole to be regular at the center, by using Eq.~(\ref{eqn:freepara}), the charge $Q$ and the mass $M$ of the black hole and those of electron are related by
\be\label{eqn:bhelecrel}
\frac{81 \pi ^4 Q^6}{131072 M^4}=\frac{e^6}{45\pi\, \alpha\, m_e^4}.
\ee
One finds from this equation that an object
\footnote{This object is not, however, an electron as electrons have spin which we are not considering in our example.}
of mass $M=m_e$ and charge $Q=e$ cannot be taken as a solution to the Einstein-NED theory presented in this paper, because for these values Eq.~(\ref{eqn:bhelecrel}) is not satisfied.

We recall that in the $\mathcal{P}$ framework, there are different Lagrangians in different parts of the spacetime. Approaching the center of the black hole, $\mathcal{P}$ goes to $-\infty$. In this region the structure function (\ref{eqn:myh}) can be expressed, approximately, by $\mathcal{H}(\mathcal{P}\rightarrow -\infty)\approx\kappa^{-2}\mathcal{P}^{-1}$. Now, by using Eq.~(\ref{eqn:lfromh}), and the relation $\mathcal{F}=\mathcal{H}_{\mathcal{P}}^2\mathcal{P}\approx \kappa^{-4}\mathcal{P}^{-3}$, one finds that, near the center of the black hole, the electrodynamic Lagrangian is
\be\label{eqn:near:lag}
\mathcal{L}\approx -\frac{3\mathcal{F}^{\frac{1}{3}}}{\kappa^{\frac{2}{3}}},
\ee
which is, of course, totally non-Maxwellian. Therefore, the assumption of the no-go theorem which forbids the existence of electrically charged regular black hole in NED with Maxwellian weak field limit, is not satisfied, allowing the non-singular behavior at all $r$.

Magnetically charged black holes, however, are not covered by the no-go theorem and one can find such black hole solutions in general relativity which are regular at the center~\cite{bronnikov2001}. An adequate example is the solution found in Ref.~\refcite{kruglov2017}. In that paper a Lagrangian of NED has been introduced which the electric part of its Taylor expansion is similar to the Lagrangian (\ref{eqn:helag}).
\footnote{If one takes the parameter $\beta$ of Ref.~\refcite{kruglov2017} to be $\beta=\gamma/4$, then the electric part of the Lagrangian of Ref.~\refcite{kruglov2017} is the same as Eq.~(\ref{eqn:helag}) to second order in $\mathcal{F}$. One only needs to consider the sign convention used in the action of our paper and Ref.~\refcite{kruglov2017}.}
This Lagrangian holds all over the spacetime. One way to see it is from the plot of electric field which is a monotonic decreasing function of the radial coordinate \cite{kruglov2017}. Recall from Fig.~\ref{fig:elec_field} that in our model the electric field has two extrema. In fact, our model reduces to the Lagrangian (\ref{eqn:helag}) only at large radial coordinate (and in the weak field limit). In small radial coordinate our model reduces to the Lagrangian (\ref{eqn:near:lag}) and makes it possible for us to construct electrically charged regular black hole solutions.

\section{Concluding remarks}\label{sec:con}

We have presented the regular static, spherically symmetric electrically charged solutions of general relativity with a new theory of NED as the source. We have studied the basic thermodynamic properties of the black hole solution. The thermodynamic features are qualitatively similar to those of Reissner-Nordstr\"{o}m black holes, and include a critical point if embedded in anti de Sitter background. Further investigations of this point, like those presented in Refs.~\refcite{kubizvnak2012,Poshteh:2013pba,Poshteh:2015ova}, are left for some future studies.

The theory we have presented in this paper violates WEC, DEC, and SEC, only near the center of the solution, and, if the solution has an event horizon, it will shield the region in which the energy conditions are violated. We would like to recall here that the energy conditions are not some physical constraints per se, but rather some mathematical boundary conditions imposed to limit the solutions of a theory. In fact, there are plethora of results that show energy conditions may/must be violated, theoretically or according to observations; as one sees that physically traversable wormholes violate WEC~\cite{morris1988}, or a positive cosmological constant violates SEC~\cite{hawking1973}. On the other hand, energy conditions are known to be violated by quantum fields~\cite{casimir1948}. Meanwhile, in the theory proposed in this paper, WEC, DEC, and SEC are all satisfied outside the event horizon of the black hole.

At far distances from this static, spherically symmetric black hole, the structure function introduced in this paper, leads to a Lagrangian which corresponds to Heisenberg-Euler Lagrangian of quantum electrodynamics. Only, it is needed that the mass and the charge of the black hole obey Eq.~(\ref{eqn:bhelecrel}). Some comments on this equation are in order.

\textit{First}, plugging in the values of the charge and the mass of electron and using the metric function, we find that the event horizon could only exist for $M \gtrsim 2.26269\times 10^7\,{\rm m}$ and $\Lambda\ll M^{-2}$ (note that the mass of an intermediate-mass black hole is in the range $10^5-10^8\,{\rm m}$). The \textit{second} point one reads from Eq.~(\ref{eqn:bhelecrel}) is that, because electric charge is a quantized quantity, so must be the mass $M$ of the black hole.
We recall that these results are true if the mass of the black hole results only from the electromagnetic field source which is described by the structure function \eqref{eqn:myh}. Also, this black hole is a static, spherically symmetric solution of general relativity with this source. It remains to be checked if these results are also true for nonstatic solutions.

\section*{Acknowledgments}

We thank Kirill Bronnikov for useful comments.

\bibliographystyle{ws-ijmpd}
\bibliography{sample}
\end{document}